\documentclass[11pt,a4paper]{article}
\usepackage{jheppub}
\usepackage{amsmath}
\usepackage{amssymb}

\title{BRST Theory for Continuous Spin}

\author[a,1]{Anders K. H. Bengtsson\note{Work supported by the Research and Education Board at the University of Bor{\aa}s.}}

\affiliation[a]{School of Engineering, University of Bor{\aa}s, All\' egatan 1, SE-50190 Bor\aa s, Sweden.}

\emailAdd{anders.bengtsson@hb.se} 

\abstract{Some puzzling aspects of higher spin field theory in Minkowski space-time, such as the tracelessness constraints and the search for an underlying physical principle, are discussed. A connecting idea might be provided by the recently much researched continuous spin representations of the Poincar{\'e} group. The Wigner equations, treated as first class constraints, yields to a four-constraint BRST formulation. The resulting field theory, generalizing free higher spin field theory, is one among a set of higher spin theories that can be related to previous work on unconstrained formulations. In particular, it is conjectured that the unconstrained higher spin theory of Francia and Sagnotti is a limit of a continuous spin theory. Furthermore, a simple analysis of the constraint structure reveals a hint of a physical rationale behind the trace constraints.}


\begin{document}

\maketitle

\pagebreak

\section{Introduction}\label{sec:Introduction}
A most mysterious, and awkward, aspect of higher spin gauge field theory is the double tracelessness and tracelessness constraints for fields $\varphi_{\mu_1\ldots\mu_s}$ gauge parameters $\theta_{\mu_1\ldots\mu_{s-1}}$ respectively
\begin{align*}
\varphi^{\alpha\beta}_{\,\,\,\,\,\,\alpha\beta\mu_{5}\ldots\mu_{s}}&=0\quad\quad\text{for}\quad s\geq4,\\
\theta^\alpha_{\,\,\,\alpha\mu_{3}\ldots\mu_{s-1}}&=0\quad\quad\text{for}\quad s\geq3.
\end{align*}
Their presence, and consequently their need to be treated, is a nuisance. Superficially, and technically, it is since long well understood why they appear. There is a mismatch between the number of physical degrees of freedom of the higher spin massless states, and the number of components of the corresponding tensor gauge fields. This mismatch cannot be accounted for by ordinary gauge transformations only. Instead, these extra trace and double trace constraints are imposed on the gauge parameters and the gauge fields respectively. This is all well known since the canonical papers \cite{Fronsdal1978,deWitFreedman} on higher spin gauge fields. 

In the original BRST treatment of higher spin gauge field theory the trace constraints were treated as second class and imposed as equations on the states \cite{OuvryStern1987a,AKHB1986a}. Alternative formulations were proposed about ten years ago, employing non-local actions \cite{FranciaSagnotti2002a,FranciaSagnotti2003a} or extra compensator fields \cite{FranciaSagnotti2005a,FranciaMouradSagnotti2007a,SagnottiTsulaia2004a,BekaertBoulanger2003}\footnote{For a more extensive set or references as well as a thorough discussion, see \cite{CampoleoniFranciaMouradSagnotti2009}.}. These latter approaches essentially amount to introducing new degrees of freedom in order that the concomitant extra gauge invariance alone can fix the mismatch in numbers of degrees of freedom. Still, one cannot escape the impression that something is missing in the basic understanding.

There is the related problem of understanding the underlying physical principle, if any such exists, behind higher spin gauge fields. In \cite{AKHB2009a}, mechanical models were briefly discussed for higher spin gauge fields as a step towards basing a physical picture of the interactions on such a model. Although looking promising at the outset, such an approach is fraught with problems that presumably go back to the free theory itself and its constraint structure. In \cite{AKHB1988} and \cite{AKHB2007a} a tentative role for the tracelessness constraint (or a similar condition) in relation to the cubic vertex were noted.

The recent interest in the continuous spin representations of the Poincar{\'e} group \cite{SchusterToro2103a,SchusterToro2103b,SchusterToro2103c} now seems to offer a hope for finding the missing idea that might connect these loose ends. Indeed, the four Wigner equations \cite{BargmannWigner1947} for the wave function, can be interpreted as first class constraints of a mechanical model. The resulting BRST theory is the subject of the present paper. Sections \ref{sec:ConstraintStructure} through \ref{sec:FourConstraintTheory} sets up the background for the BRST treatment of section \ref{sec:BRSTTheory}. The tentative underlying physics is discussed in section \ref{sec:PhysicalRationale}. Some concluding remarks are in section \ref{sec:ConcludingRemarks} and some technical details about the ghost complexes are relegated to section \ref{sec:NotationConventions}.

\section{Constraint structure}\label{sec:ConstraintStructure}
In a BRST approach to free higher spin gauge fields, the gauge transformations are generated by the first class constraints of an underlying mechanics model, while the trace conditions can be imposed through second class constraints. This works well for the free theory and reproduces the Fronsdal theory. Let us work in four dimensions (with a $-+++$ metric) and briefly iterate the steps.

We start with a classical or first-quantized two-particle relativistic mechanical system with center of motion $(x_\mu,p_\nu)$ and relative $(\xi_\mu,\pi_\nu)$ coordinates and momenta, but we don't specify any action, instead working directly from the constraints. For the relative coordinates we also use oscillators $(\alpha_\mu,\alpha^\dagger_\nu)$, or holomorphic coordinates classically, defined in terms of the relative coordinates and momenta as
\begin{equation}\label{eq:OscillatorTranscription}
\alpha_\mu=\frac{1}{\sqrt 2}(\xi_\mu+i\pi_\mu),\quad\alpha_\mu^\dagger=\frac{1}{\sqrt 2}(\xi_\mu-i\pi_\mu).
\end{equation}
We take $\xi_\mu$ and $\pi_\nu$ to be dimensionless. Classically we have
\begin{align}\label{eq:ClassicalBrackets}
\{x_\mu,p_\nu\}=\eta_{\mu\nu},\quad\{\xi_\mu,\pi_\nu\}=\eta_{\mu\nu},
\end{align}
and quantum mechanically
\begin{align}\label{eq:QuantumBrackets}
[x_\mu,p_\nu]=i\eta_{\mu\nu},\quad [\xi_\mu,\pi_\nu]=i\eta_{\mu\nu},\quad [\alpha_\mu,\alpha^\dagger_\nu]=\eta_{\mu\nu}.
\end{align}
Excluding explicit occurrence of the center of motion coordinate $x_\mu$ there are six bilinear scalars in terms of these variables
\begin{align}\label{eq:ConstraintsFirstClass}
G_0&=-{1\over 2}p^2,\quad G_+=\alpha\cdot p,\quad\quad\, G_-=\alpha^\dagger\cdot p,\\\label{eq:ConstraintsSecondClass}
T&={1\over 2}\alpha\cdot\alpha,\quad T^\dagger={1\over 2}\alpha^\dagger\cdot\alpha^\dagger,\quad N={1\over 2}(\alpha\cdot\alpha^\dagger+\alpha^\dagger\cdot\alpha)=\alpha^\dagger\cdot\alpha+2.
\end{align}
From this set we can chose various linear combinations as first and second class constraints by (weakly) equating to zero. Once such a choice is made, ghost coordinates and momenta can be introduced corresponding to the first class set and the BRST operator $Q$ constructed. Then a free field theory can be set up using BRST-BV techniques. The standard choice is simply taking the set $\{G_0=0\,,\,G_+=0\,,\,G_-=0\}$ as first class. Then the tracelessness constraints (on fields and parameters) are given by $T|\mathrm{state}\rangle=0$ with the $T$ operator augmented with a ghost contribution.

One obvious problem with this approach is that it is very formal, inspired as it is (and initially was) by string field theory \cite{Siegel1984a,West1986SlacPp,BanksPeskin1986}. Where, is the physics? For strings, there is vibration dynamics that motivates the introduction of oscillators. For higher spins, such dynamics is not at all obvious. There are various possible limits of strings (such as the much researched zero-tension limit\footnote{This limit has been studied by many authors throughout the history of string theory.}, the straight line (or rigid) string  \cite{CasalbuoniLonghi1975} and the discrete string \cite{GershunPashnev1988}) that furnish constraints that are essentially built from different linear combinations of the bilinear terms \eqref{eq:ConstraintsFirstClass} and \eqref{eq:ConstraintsSecondClass}. But the physical intuition is weak. A nice picture is a one-dimensional ''spring model'' of two relativistic particles bound by a harmonic potential. Such vibrations could correspond to a constraint $\pi^2+\xi^2=0$. However, this constraint, corresponding to the oscillator Hamiltonian, would fix the excitation level and reduce the field theory to a single spin theory, whereas we want to accommodate all spins. Clearly, almost any sensible two-particle action with symmetries will produce constraints that are various linear combinations of the set \eqref{eq:ConstraintsFirstClass} and \eqref{eq:ConstraintsSecondClass}. For a thorough discussion, see reference \cite{HenneauxTeitelboim1989a}.

\section{Continuous spin}\label{sec:ContinuousSpin}
Now for some input from continuous spin theory.\footnote{The theory goes back to Wigner \cite{Wigner1939}. A set of classic references are \cite{Yngvason1970,IversonMack1971,Hirata1977,Zoller1994}. The recent literature include \cite{BrinkKhanRamondXiong2002,EdgrenMarneliusSalomonsson2005,EdgrenMarnelius2006,BekaertMourad2006}.} In the Bargmann-Wigner paper \cite{BargmannWigner1947} the equations are given as
\begin{align}\label{eq:EdgrenEtAlConstraints}
p^2\psi&=0,\\
\xi^2\psi&=-l\psi,\\
p\cdot\xi\psi&=0,\\
p\cdot\partial/\partial\xi\psi&=-i\Xi\psi.
\end{align}
where $\psi(p,\xi)$ is a wave function, $p$ the momentum operator and $\xi$ an internal four-vector of length $l$ ($l=1$ in \cite{BargmannWigner1947}). These equations look conspicuously much like constraint equations for higher spin gauge fields. To make the connection more precise, let us instead follow \cite{EdgrenMarneliusSalomonsson2005,EdgrenMarnelius2006}. These authors consider the equations
\begin{align}\label{eq:EdgrenEtAlPhys}
p^2|\rm{phys}\rangle=0,\\
(w^2-\mu^2)|\rm{phys}\rangle=0.
\end{align}
to be satisfied by physical states $|\rm{phys}\rangle$ and where $w^\mu$ is the Pauli-Lubanski vector
\begin{equation}\label{eq:PauliLubanski}
w^\mu=\frac{1}{2}\epsilon^{\mu\nu\rho\sigma}m_{\nu\rho}p_\sigma.
\end{equation}
They find the most general solution in terms of four constraints. We will not need the most general case, so let us just pick the set (using slightly different conventions)
\begin{align}\label{eq:CSConstraintsXiPi0}
G_0&=-\frac{1}{2}p^2,\\\label{eq:CSConstraintsXiPi1}
G_1&=p\cdot\pi,\\\label{eq:CSConstraintsXiPi2}
G_2&=p\cdot\xi-\frac{\mu}{\omega},\\\label{eq:CSConstraintsXiPi3}
G_3&=-\pi^2+\omega^2.
\end{align}
These are essentially the same equations as considered in \cite{BekaertMourad2006}. Classically, we can think of the equations as set of four first class constraints $G_i\approx0$. The non-zero Poisson brackets are
\begin{equation}\label{eq:CSAlgebraXiPi}
\{G_1,G_2\}=-2G_0,\quad\quad \{G_2,G_3\}=-2G_1.
\end{equation}
With $\mu$ zero and neglecting $G_3$ these are precisely the first class constraints that upon BRST-quantization yield a free field theory containing all integer spin and where the field equations are precisely those of Fronsdal after imposing the trace constraint. As will argued in the next section, the continuous spin constraint $G_3$ can serve a role similar to the tracelessness constraint.

\section{Four-constraint theory}\label{sec:FourConstraintTheory}
Before using the input from continuous spin representations, let us first see what can be done with the constraint structure from a purely formal point of view. The idea is to treat the trace constraint as first class. Counting phase space degrees of freedom (d.o.f) as $(\#\text{coordinates},\#\text{momenta)}$ for center of motion (cm.) and relative motion (rel.) and adding, we have in four dimensions: $(4,4)_{\rm{cm.}}+(4,4)_{\rm{rel.}}$ that is $16$ d.o.f. in all. The three first class constraints $\{G_0=0\,,\,G_+=0\,,\,G_-=0\}$ remove $3\cdot2=6$ d.o.f. The two second class constraints $\{T=0\,,\,T^\dagger=0\}$ remove $2$ d.o.f. We are left with $8$ d.o.f. which in phase space distribute as $(3,3)_{\rm{cm.}}+(1,1)_{\rm{rel.}}$. This precisely corresponds to a tower of massless fields with helicities $\pm \lambda$. There is a simple way to treat a pair of second class constraints as first class: take one of them as a first class constraint and treat the remaining one as a gauge condition. Two requirements must be met: the new constraint algebra must be first class and the new constraint must be Hermitean (real).

Then it is clear that we cannot take $T=0$ as first class since it is not Hermitean. Here we need not run through the full analysis, suffice it to say that any linear combination of the set $\{T,T^\dagger,N\}$ that is Hermitean would do. We will study three cases in this paper, see equation \eqref{eq:LinearCombinationG3} below.

Let us now be specific and pick $G_3$ from equation \eqref{eq:CSConstraintsXiPi3} as the fourth first class constraint. Using \eqref{eq:OscillatorTranscription} we find
\begin{equation}\label{eq:CSFourthConstraint}
G_3=-\pi^2+\omega^2=T+T^\dagger-N+\omega^2.
\end{equation}
We keep $G_0$ as it is, but linearly recombine $G_1$ and $G_2$ into
\begin{align}\label{eq:CSGplusGminusConstraints}
G_+&=p\cdot\alpha-\frac{\mu}{\sqrt{2} \omega},\\
G_-&=p\cdot\alpha^\dagger-\frac{\mu}{\sqrt{2} \omega},
\end{align}
so that
\begin{equation}\label{eq:ConstraintLinCombs}
G_++G_-=\sqrt{2}G_2,\quad\quad G_+-G_-=i\sqrt{2}G_1.
\end{equation}
The non-zero commutators of the resulting first class algebra is
\begin{align}\label{eq:CSAlgebraOscillator}
[G_+,G_-]&=-2G_0,\\
[G_+,G_3]&=G_- - G_+,\\
[G_-,G_3]&=G_- - G_+.
\end{align}
In the following, $\mu/\sqrt2 \omega$ will be denoted by $\beta$.
\subsection*{A slight generalization}
It is interesting to see whether the $G_3$ constraint can be generalized since this is really where we depart from standard BRST higher spin. Instead of $G_3$ we may attempt a constraint $G_4$ with the following non-zero commutators with the set $\{G_0,G_+,G_-\}$
\begin{align}\label{eq:G4Algebra1}
[G_+,G_4]&=\sigma_- G_- +\sigma_+ G_+,\\\label{eq:G4Algebra2}
[G_-,G_4]&=-\sigma_+ G_- -\sigma_-G_+.
\end{align}
Taking $G_4=N$ corresponds to $\sigma_+=1$ and $\sigma_-=0$ but then we must have $\beta=0$. Likewise taking $G_4=T+T^\dagger$ corresponds to $\sigma_+=0$ and $\sigma_-=1$ and again we must have $\beta=0$. The case $G_4=G_3$ corresponds to $\sigma_-=-\sigma_+=1$ and then we can have $\beta\not=0$. This is the continuous spin case. The different cases can be captured by writing 
\begin{equation}\label{eq:LinearCombinationG3}
G_3=\sigma_-\big(T+T^\dagger\big)+\sigma_+N+\omega^2
\end{equation}
where from now on we use the notation $G_3$ for this generalized case. Taking $\sigma_+=\sigma_-=1$ would correspond to taking $G_3=\xi^2+\omega^2$, but since $\pi$ and $\xi$ can be interchanged through canonical transformations \cite{EdgrenMarnelius2006}, nothing new is gained by this.

For clarity, the interesting cases are summarized in the Table \ref{tab:TheoriesConsidered}: Higher spin with ''trace constraint'' $T+T^\dagger=0$ (abbreviated $\mathrm{HS}_T$), Higher spin with ''number constraint'' $N=0$ (abbreviated $\mathrm{HS}_N$) and Continuous spin (abbreviated $\mathrm{CS}$).
\begin{table}[h]
\centering
\begin{tabular}{l|c|c|l}
Type of theory & $\sigma_+$ & $\sigma_-$ & Comment\\\hline
$\mathrm{HS}_T$ & $0$ & $1$ & requires $\beta=0$ but allows $\omega^2\not=0$\\
$\mathrm{HS}_N$ & $1$ & $0$ & requires $\beta=0$ but allows $\omega^2\not=0$\\
$\mathrm{CS}$ & $-1$ & $1$ & $\beta\not=0$, $\omega^2\not=0$ (but allows $\beta=0$, $\omega^2=0$)\\ 
\end{tabular} 
\caption{Theories considered.}
\label{tab:TheoriesConsidered}
\end{table}

A four-constraint theory of the type $\mathrm{HS}_N$ has been studied previously in \cite{Meurice1988}.

\section{BRST theory}\label{sec:BRSTTheory}
The free field theory of massless higher spin fields has been extensively studied by many authors during the last decade (for review and references, see \cite{CampoleoniFranciaMouradSagnotti2009}). Much effort has been spent in trying to circumvent the tracelessness constraints, one interesting approach being the unconstrained formulation using compensator fields originally proposed in \cite{FranciaSagnotti2002a,FranciaSagnotti2003a}. In section \ref{subsec:FranciaSagnotti} we will see how that formulation can be understood in the present framework.

\subsection{Backdrop to higher spin}
Without $G_3$ and with $\beta=0$, we would have the higher spin BRST-operator $Q_{hs}$ 
\begin{equation}\label{eq:Qhs}
Q_{hs}=c^0G_0+c^+G_++c^-G_-+2c^+c^-b_0.
\end{equation}
For ghost and vacuum conventions, see section \ref{sec:NotationConventions}. The fields and gauge parameters
are expanded as
\begin{align}\label{eq:HSFieldExpansion}
|\Phi\rangle&=(A+Fc^+b_-+Hb_-c^0)|+\rangle,\\
|\Theta\rangle&=\theta b_-|+\rangle.
\end{align}
Here $A$ contains the symmetric integer spin tensors, $H$ are auxiliary fields and $F$ contains the traces of the fields in $A$ upon imposing the trace constraint. All fields are expanded over the oscillator basis. For the complete field and anti-field BV-complex, the reader is referred to \cite{AKHB2007a}. Field equations and gauge transformations are given by
\begin{align}\label{eq:HSFieldEqGaugeTrans}
Q_{hs}|\Phi\rangle&=0,\\
\delta|\Phi\rangle&=Q_{hs}|\Theta\rangle.
\end{align}
The content of the field equations can be extracted as
\begin{align}\label{eq:HSFieldEquations1}
\big(G_0A+G_-H\big)|-\rangle=0,\\\label{eq:HSFieldEquations2}
\big(G_0F+G_+H\big)c^+b_-|-\rangle=0,\\\label{eq:HSFieldEquations3}
\big(G_+A-G_-F-2H\big)c^+|+\rangle=0.
\end{align}
These equations are sometimes called the ''triplet'' equations \cite{FranciaSagnotti2003a,FranciaSagnotti2005a}. The last equation can be solved algebraically for the auxiliary field $H$. Then applying the trace constraint $\big(T+b_+c^-\big)|\Phi\rangle=0$, traces of the $A$ fields are related to the $F$ fields through $A=TF$. The standard Fronsdal equations result for the integer spin field components in $A$. Indeed, if all the fields are expanded over the oscillator basis, generically as
\begin{equation}\label{eq:OscillatorExpansionField}
\phi=\phi_0+\phi_1^\mu\alpha^\dagger_\mu+\phi_2^{\mu\nu}\alpha^\dagger_\mu\alpha^\dagger_\nu+\cdots ,
\end{equation}
and the shorthand notation $A^{(s)}$ for a gauge field $A_{\mu_1\mu_2\ldots\mu_s}$ with $s$ symmetrized indices is used, equations \eqref{eq:HSFieldEquations1} and \eqref{eq:HSFieldEquations3} yield 
\begin{equation}\label{eq:HSFieldEquation4}
p^2A^{(s)}-p^{(1}p\cdot A^{s)}+p^{(1}p^{2}F^{s-2)}=0,
\end{equation}
with $p=-i\partial$. The trace constraint finally gives $F^{(s-2)}=A^{\prime(s)}$ (where the prime denotes a trace), thus recovering the Fronsdal equations. However, the field equation \eqref{eq:HSFieldEquation4} for the doublet $(A,F)$ is gauge invariant even without imposing the trace constraint, and thus is unconstrained.

It can be noted that this is all true even if $\beta\not=0$ in the constraints $G_+$ and $G_-$. A non-zero $\beta$ would give level, i.e. spin, mixing, a phenomena that will also result in the continuous spin case. 

\subsection{Continuous spin}
With this background, let us return to the four-constraint theory. Since the fourth constraint is Hermitean, the fourth ghost pair $(c^3,b_3)$ must also be Hermitean. Then the continuous spin BRST-operator $Q_{cs}$ is
\begin{equation}\label{eq:Qcs}
Q_{cs}=Q_{hs}+c^3G_3-c^3(c^+ + c^-)(b_+ - b_-).
\end{equation}
The slightly more general form is
\begin{equation}\label{eq:Qcs}
Q_{gen}=Q_{hs}+c^3G_3+\sigma_+c^3(b_-c^--b_+c^+)+\sigma_-c^3(b_+c^--b_-c^+).
\end{equation}
The interpretation of the extra terms is clear. When $\sigma_+=1$ and $\sigma_-=0$, the extra term is a ghost number operator. On the other hand,  when $\sigma_+=0$ and $\sigma_-=1$, the extra term is a ghost trace operator.

The presence of a second Hermitean ghost pair leads to a second vacuum degeneracy above the degeneracy between $|+\rangle$ and $|-\rangle$. The vacuum structure becomes somewhat more complicated. Using the notation $|\pm_0\pm_3\rangle$ so that the first entry refers to the $0$\,-ghosts and the second to the $3$\,-ghosts, the vacuum complex is given in Table \ref{tab:CSVacuumComplex}.

\begin{table}[h]
\centering
\begin{tabular}{ccccc}
 & $0$ & & $0$ &\\
 & ${\uparrow\atop b_3}$ & & ${\uparrow\atop b_3}$ &\\
$0\,\,{\atop \overleftarrow{b_0}}$ & $|++\rangle$ & ${\overrightarrow{c^0}\atop \overleftarrow{b_0}}$ & $|-+\rangle$ & $ {\atop \overrightarrow{c^0}}\,\,0$\\

 & ${c^3\atop\downarrow}{\uparrow\atop b_3}$ & & ${c^3\atop\downarrow}{\uparrow\atop b_3}$ &\\

$0\,\,{\atop \overleftarrow{b_0}}$ & $|+-\rangle$ & ${\overrightarrow{c^0}\atop \overleftarrow{b_0}}$ & $|--\rangle$ & $ {\atop \overrightarrow{c^0}}\,\,0$\\
& ${c^3\atop\downarrow}$ & & ${c^3\atop\downarrow}$ &\\
 & $0$ & & $0$ &
\end{tabular}
\caption{Doubly degenerate vacuum complex.}
\label{tab:CSVacuumComplex}
\end{table}

This complex is built from the generic principle 
\begin{equation}\label{eq:GenericGhostVacua}
0\,\,\overleftarrow{b} \;\;|+\rangle\;\; {\overrightarrow{c}\atop \overleftarrow{b}} \;\;|-\rangle\;\; \overrightarrow{c}\,\,0.
\end{equation}
As for Grassmann parities, it is consistent to chose $|++\rangle$ and $|--\rangle$ to have parity $0$ and $|+-\rangle$ and $|-+\rangle$ to have parity $1$. The full ghost complex is given in section \ref{sec:NotationConventions} in Table \ref{tab:FourConstraintSpinGhostComplex}.

The presence of the two Hermitean ghost pairs makes it difficult, perhaps impossible, to set up a BRST Lagrangian of the type $\langle\Phi |Q|\Phi\rangle$ (or alternatively a BV master action $S$) in any simple way \cite{LyakhovichSharapov2004}. Therefore we will recourse to field equations for the time being\footnote{This question is further discussed in section \ref{subsec:FranciaSagnotti}.}. The physical fields reside in the (mechanical) ghost number $-1$ sector and the gauge parameters in the ghost number $-2$ sector
\begin{align}\label{eq:CSFieldExpansion}
|\Phi\rangle&=(A+Fc^+b_-+Hb_-c^0+Bb_-c^3)|++\rangle,\\\label{eq:CSParamExpansion}
|\Theta\rangle&=\theta b_-|++\rangle,
\end{align}
where $B$ is the new field corresponding to the trace constraint.

The field equations following from $Q_{cs}|\Phi\rangle=0$ now become
\begin{align}\label{eq:CSFieldEquations1}
\big(G_0A+G_-H\big)|-+\rangle=0,\\\label{eq:CSFieldEquations2}
\big(G_0F+G_+H\big)c^+b_-|-+\rangle=0,\\\label{eq:CSFieldEquations3}
\big(G_+A-G_-F-2H\big)c^+|++\rangle=0,\\\label{eq:CSFieldEquations4}
\big(G_0B-G_3H\big)b_-|--\rangle=0,\\\label{eq:CSFieldEquations5}
\big(G_+B+G_3F+A-F\big)c^+b_-|+-\rangle=0,\\\label{eq:CSFieldEquations6}
\big(G_-B+G_3A+A-F\big)|+-\rangle=0.
\end{align}
Here, the first three equations are the same as in higher spin theory. The last two equations can also be written more generally, corresponding to \eqref{eq:G4Algebra1} and \eqref{eq:G4Algebra2}, as
\begin{align}\label{eq:CSFieldEquationsGen5}
\big(G_+B+G_3F+\sigma_-A+\sigma_+F\big)c^+b_-|+-\rangle=0,\\\label{eq:CSFieldEquationsGen6}
\big(G_-B+G_3A-\sigma_+A-\sigma_-F\big)|+-\rangle=0.
\end{align}
Working with these equations, we can compare different constraint structures.\footnote{A theory with four constraints was studied in the paper \cite{BuchbinderGalajinskyKrykhtin2007}. These authors treat the second class constraints in a different way.}

\subsection{The field equations}\label{subsec:FieldEquations}
Since the starting point for this investigation was the puzzle of the tracelessness constraints, let us begin with how they come about in the present theory by analyzing the field equations. From now on the ghosts and the vacua are dropped from the notation but all equations should be thought of as acting on a vacuum $|\mathrm{vac}\rangle$ for the bosonic oscillators. 

Using the field equation \eqref{eq:CSFieldEquationsGen6}, we can solve for the field $F$ in the form
\begin{equation}\label{eq:FTraceEqCS}
\sigma_-F=\big(G_3-\sigma_+\big)A+G_-B,
\end{equation}
and since $G_3$ computes the trace, among other things, this equation will allow us to derive a variant of the higher spin equations. Writing $G_3=T+\tilde{G_3}$ where $\tilde{G_3}=T^\dagger-N+\omega^2$, equation \eqref{eq:FTraceEqCS} allows us to write the higher-spin similar field equation as
\begin{equation}\label{eq:CSHSSimilarFieldEquation}
\begin{split}
&p^2A-G_-G_+A+\frac{1}{\sigma_-}G_-G_-TA=\\&-\frac{1}{\sigma_-}G_-G_-\big(T^\dagger-N+\omega^2-\sigma_+\big)A-\frac{1}{\sigma_-}G_-G_-G_-B.
\end{split}
\end{equation}
The left hand side corresponds to Fronsdal's equations, although due to the special form of the CS constraints $G_-=p\cdot\alpha^\dagger-\beta$ and $G_+=p\cdot\alpha-\beta$ there are extra contributions, of order $\beta$ and $\beta^2$ mixing fields from various excitation levels. For instance, the field equation for the spin $0$ field $A^{(0)}$ will contain terms $\sim\beta p\cdot A^{(1)}$ and $\sim\beta^2 A^{\prime(2)}$. This was found in the recent paper \cite{SchusterToro2103c}. The phenomena repeats itself on the spin $1$ level where there occur derivatives of the spin $2$ field and the trace of the spin $3$ field.

The right hand side of \eqref{eq:CSHSSimilarFieldEquation} provides source terms. It is zero for spin $1$. For the spin $2$ field $A_2$, it provides a source term of the form
\begin{equation}
-\frac{\omega^2-\sigma_+}{\sigma_-}p_{\mu}p_{\nu}A_0.
\end{equation}
This term is not present in the $\mathrm{HS}_T$ theory (see Table \ref{tab:TheoriesConsidered}). The last term on the right hand side starts to contribute source terms for the spin $3$ field $A^{(3)}$ where it produces a term of the form $p_{\mu_1}p_{\mu_2}p_{\mu_3}B_0$. The term $G_-G_-G_-B$ corresponds to the compensator field of \cite{FranciaSagnotti2005a,FranciaSagnotti2006rw} first appearing at the spin $3$ level. For more comments on this, see section \ref{subsec:FranciaSagnotti}.

However, equation \eqref{eq:CSFieldEquationsGen5} have a very similar structure and it allows us to write
\begin{equation}\label{eq:ATraceEqCS}
\sigma_-A=-\big(G_3+\sigma_+\big)F-G_+B.
\end{equation}
The structure is thus quite complex and working spin level by spin level soon becomes confusing. In order to analyze the full content of the equations let us arrange them in three groups. 
\begin{description}
  \item[Dynamical equations] 
\begin{align}\label{eq:DynFieldEquation1}
G_0A+G_-H=0,\\\label{eq:DynFieldEquation2}
G_0F+G_+H=0,\\\label{eq:DynFieldEquation3}
G_0B-G_3H=0,
\end{align}

  \item[Auxiliary equation]
\begin{align}\label{eq:AuxFieldEquation}
H=\frac{1}{2}\big(G_+A-G_-F\big),
\end{align}

  \item[Trace equations]
\begin{align}\label{eq:TraceFieldEquationF}
\sigma_-F&=\big(G_3-\sigma_+\big)A+G_-B,\\\label{eq:TraceFieldEquationA}
\sigma_-A&=-\big(G_3+\sigma_+\big)F-G_+B.
\end{align}
\end{description}
In principle, we do not expect to have six independent field equations for four different fields, so there must be redundancies in these equations. Since the trace equations are first order in derivatives they should really be counted as one equation. This leaves one equation to many. In section \ref{subsec:FormalSolutionFieldEquations} we will see that equation \eqref{eq:DynFieldEquation3} can either be seen as a trivial identity or as a consequence of the trace equations. The effective number of field equations are therefore just four.

\subsection{Gauge transformations and invariance}\label{subsec:GaugeTransfromationsInvariance}
With the field and gauge parameter expanded over the ghost complex as in \eqref{eq:CSFieldExpansion} and \eqref{eq:CSParamExpansion} the gauge transformations become
\begin{align}
\delta A&=G_-\theta,\\
\delta F&=G_+\theta,\\
\delta H&=-G_0\theta,\\
\delta B&=-G_3\theta.
\end{align}
The invariance of the field equations \eqref{eq:DynFieldEquation1} to \eqref{eq:TraceFieldEquationA} are direct consequences of the gauge constraint algebra. 

\subsection{Formal solution of the field equations}\label{subsec:FormalSolutionFieldEquations}
The structure of the field equations suggest two formal ways of reducing them.

\subsubsection*{Solving for $A$ and $F$}

The two fields $A$ and $F$ are coupled through the auxiliary field $H$. Inserting $H$ from \eqref{eq:AuxFieldEquation} into \eqref{eq:DynFieldEquation1} and \eqref{eq:DynFieldEquation2} and collecting the fields into a vector, we get
\begin{align}
G_0\begin{bmatrix}A \\ F\end{bmatrix}
+\frac{1}{2}\begin{bmatrix}G_-G_+ & -G_-G_- \\ G_+G_+ & -G_+G_-\end{bmatrix}
\begin{bmatrix}A \\ F\end{bmatrix}=0.
\end{align}
Then the trace equations \eqref{eq:TraceFieldEquationF} and \eqref{eq:TraceFieldEquationA} can be written in matrix form as 
\begin{equation}\label{eq:MatrixTraceFieldEquations}
\mathbf{T}
\begin{bmatrix}
A \\ F
\end{bmatrix}
=
-\begin{bmatrix}
G_- \\ G_+
\end{bmatrix}
B,
\end{equation}
where the matrix $\mathbf{T}$ is
\begin{equation}\label{eq:TraceMatric}
\mathbf{T}=
\begin{bmatrix}
G_3 - \sigma_+ & -\sigma_-\\ \sigma_- & G_3 + \sigma_+
\end{bmatrix}.
\end{equation}
It is formally invertible with inverse
\begin{equation}\label{eq:InverseTraceMatric}
\mathbf{T}^{-1}=\frac{1}{\sigma_-^2-\sigma_+^2+G_3^2}
\begin{bmatrix}
 G_3 + \sigma_+ & \sigma_- \\ -\sigma_-  &  G_3 - \sigma_+
\end{bmatrix}.
\end{equation}
This means that the doublet of fields $(A,F)$ is expressible in terms of the $B$ field as
\begin{equation}\label{eq:AFintermsofB}
\begin{bmatrix}
A \\ F
\end{bmatrix}
=-\mathbf{T}^{-1}
\begin{bmatrix}
G_- \\ G_+
\end{bmatrix}
B.
\end{equation}
The inverse $\mathbf{T}^{-1}$ is well defined in all cases. For instance in the $\mathrm{HS}_T$ case we have
\begin{equation}\label{eq:InverseTGeneral}
\mathbf{T}_{\mathrm{hs}_T}^{-1}=(1-G_3^2+G_3^4-\cdots)
\begin{bmatrix}
G_3 & 1 \\ -1 & G_3 
\end{bmatrix}.
\end{equation}
There is no issue of convergence since these operators act on fields $B|\mathrm{vac}\rangle$ and for any finite excitation level (spin level) only a finite number of terms contribute. Let us however focus on the CS case. Then we get
\begin{equation}\label{eq:InverseTCS}
\mathbf{T}_{\mathrm{cs}}^{-1}=\frac{1}{G_3^2}
\begin{bmatrix}
G_3 - 1 & 1 \\ -1 & G_3 + 1 
\end{bmatrix}.
\end{equation}
At least as long as $\omega^2\not=0$ we can expand $1/G_3^2$ in a formal power series.

Combining field equations \eqref{eq:DynFieldEquation3} and \eqref{eq:AuxFieldEquation}, the equation for $B$ can be written as
\begin{equation}\label{eq:FieldEquationBpre}
G_0B-\frac{1}{2}G_3
\begin{bmatrix}
G_+\;, &-G_-
\end{bmatrix}
\begin{bmatrix}
A \\ F
\end{bmatrix}
=0.
\end{equation}
Then using \eqref{eq:AFintermsofB} we get
\begin{equation}\label{eq:FieldEquationBfin}
G_0B+\frac{1}{2}G_3
\begin{bmatrix}
G_+\;, &-G_-
\end{bmatrix}
\mathbf{T}^{-1}
\begin{bmatrix}
G_- \\ G_+
\end{bmatrix}B=0.
\end{equation}
Thus it seems that we have reduced the content of the field equations to the equation \eqref{eq:FieldEquationBfin} for the independent field $B$, the dependent field doublet $(A,F)$ being determined by equation \eqref{eq:MatrixTraceFieldEquations}. However, equation \eqref{eq:FieldEquationBfin} is void of content. Using the gauge algebra, it can be shown that for CS
\begin{equation}\label{eq:StrangeIdentity}
\frac{1}{2}G_3
\begin{bmatrix}
G_+\;, &-G_-
\end{bmatrix}
\mathbf{T}^{-1}
\begin{bmatrix}
G_- \\ -G_+
\end{bmatrix}=-G_0,
\end{equation}
as an identity.

\subsubsection*{Solving for $B$}
Alternatively, we can view the trace equation \eqref{eq:MatrixTraceFieldEquations} as expressing $G_-B$ and $G_+B$ in terms of $A$ and $F$. Then multiplying the $G_-B$ equation by $G_+$ and the $G_+B$ equation by $G_-$ and subtracting gives
\begin{equation}\label{eq:TraceFieldEquationB1}
\big(G_+G_--G_-G_+\big)B=G_3\big(G_-F-G_+A\big),
\end{equation}
which is precisely the field equation \eqref{eq:DynFieldEquation3} with $H$ substituted for through \eqref{eq:AuxFieldEquation}. This shows that the set of field equations are compatible\footnote{Which they must be, derived as they are from a nilpotent BRST-operator based on a closed first class algebra.}, but it also shows that the field $B$ should properly be regarded as a redundant field, although it seems to have a dynamical field equation. Its field equation is however not independent, but follows from the trace equations for the doublet $(A,F)$ and the equation for $H$.

It is indeed possible to gauge $B$ to anything (but not zero). Let us see how this works out on the component level. Expand the $B$ field as in \eqref{eq:OscillatorExpansionField} and the gauge parameter as 
\begin{equation}\label{eq:OscillatorExpansionParameter}
\theta=\theta_1+\theta_2^\mu\alpha^\dagger_\mu+\theta_3^{\mu\nu}\alpha^\dagger_\mu\alpha^\dagger_\nu+\cdots ,
\end{equation}
where the indexing $\theta_s$ indicates to which primary gauge field $A^{(s)}$ the parameter belongs. Then we get for the first few levels
\begin{align*}
\delta B_0 &= -\theta_3^{\,\prime}+(2-\omega^2)\theta_1,\\
\delta B_1^{\,\mu} &= -3\theta_4^\prime + (3-\omega^2)\theta_2^{\,\mu},\\
\delta B_2^{\,\mu\nu} &= -6\theta_5^{\,\prime\mu\nu} + (4-\omega^2)\theta_3^{\,\mu\nu} - \frac{1}{2}\eta^{\,\mu\nu}\theta_0.\\
\end{align*}
What is important here is that the scalar component $B_0$ of $B$ can be gauged to anything without using up any of the freedom of the spin $1$ gauge parameter $\theta_1$. What is used is the trace part of the spin $3$ gauge parameter $\theta_3$, i.e. what is otherwise set to zero in a constrained formulation. A similar argument holds for $B_1^{\,\mu}$ and so on.

\subsection{Relation to the unconstrained formulation of Francia and Sagnotti}\label{subsec:FranciaSagnotti}
We can also make contact with the work of Francia and Sagnotti on unconstrained formulations of the higher spin field equations. For that purpose, consider the generalized case with $G_3=\sigma_-\big(T+T^\dagger\big)+\sigma_+N+\omega^2$. In their paper \cite{FranciaSagnotti2006rw} they quote the ''local non-Lagrangian compensator equations''
\begin{align}\label{eq:FranciaSagnottiEq1}
{\cal F}_{\mu_1\ldots\mu_s}&=3\partial_{\mu_1}\partial_{\mu_2}\partial_{\mu_3}\alpha_{\mu_4\ldots\mu_s}+\ldots ,\\\label{eq:FranciaSagnottiEq2}
\varphi^{\rho\sigma}_{\;\;\;\rho\sigma\mu_5\ldots\mu_s}&=4\partial\cdot\alpha_{\mu_5\ldots\mu_s}+(\partial_{\mu_5}\alpha^\rho_{\rho\mu_6\ldots\mu_s}+\ldots) ,
\end{align}
where ${\cal F}$ is a field equation for the higher spin field $\varphi$ and $\alpha$ is the compensator field. 

As we saw in equation \eqref{eq:CSHSSimilarFieldEquation}, a compensator-like term $G_-G_-G_-B$ is produced when the field $F$ is substituted for in the field equation for the field $A$. Our equation \eqref{eq:CSHSSimilarFieldEquation} would be a generalization of \eqref{eq:FranciaSagnottiEq1}. To find the generalization of \eqref{eq:FranciaSagnottiEq2}, consider the trace equations again. Substituting $F$ from \eqref{eq:TraceFieldEquationF} into \eqref{eq:TraceFieldEquationA} we get
\begin{align*}
\sigma_-^2A&=-(G_3+\sigma_+)\big((G_3-\sigma_+)A+G_-B\big)-\sigma_-G_+B\\
&=-(G_3^2-\sigma_+^2)A-(G_3+\sigma_+)G_-B-\sigma_-G_+B.
\end{align*}
Using the constraint algebra we get
\begin{equation*}
(\sigma_-^2-\sigma_+^2)A=-G_3^2A-G_-G_3B-2\sigma_+G_-B-2\sigma_-G_+B,
\end{equation*}
and precisely in the CS case when $\sigma_-^2-\sigma_+^2=0$ we get
\begin{equation}
G_3^2A=-G_-G_3B+2G_-B-2G_+B.
\end{equation}
This equation generalizes equation \eqref{eq:FranciaSagnottiEq2}. 

\subsection{Action with Lagrange multipliers}
As already remarked, it seems not possible to construct an action of the $\langle\Phi|Q|\Phi\rangle$ form for this type of four-constraint theory. One natural attempt would be to try $\langle\Phi|(c^0\pm c^3)Q|\Phi\rangle$ as this has the correct ghost number. It does not produce the correct field equations. On the level of components this shows up as a (most likely unavoidable) mixing of the dynamic equations and the trace equations.\footnote{The question needs more thought though.} It does however suggest a possible action with Lagrange multiplier fields $|\Lambda\rangle$
\begin{equation}\label{eq:LagrangeMultiplier}
|\Lambda\rangle=(\widetilde{A}+\widetilde{F}c^+b_-+\widetilde{H}b_-c^0+\widetilde{B}b_-c^3\big)|++\rangle,
\end{equation}
where we denote the components of the Lagrange multiplier fields by tildes. We take as our tentative action
\begin{equation}\label{eq:ActionLagrangeMultplier}
{\cal A}=-\frac{1}{2}\langle\Phi|c^3Q_{cs}|\Phi\rangle+\frac{\epsilon}{2}\langle\Phi|(c^0Q_{cs}-Q_{cs}c^0)|\Lambda\rangle,
\end{equation}
with $\epsilon$ parameterizing the weighting of the second term in relation to the first.\footnote{Spacetime integrations are implicit in all formulas involving actions.} The form of the second term is needed for reality since $c^0$ and $Q$ does not anti-commute. This form also ensures gauge invariance (the field $\Lambda$ does not vary under gauge transformations). Varying with respect to $|\Phi\rangle$ then produces
\begin{equation}\label{eq:VaryingPhi}
c^3Q_{cs}|\Phi\rangle-\epsilon(c^0Q_{cs}-c^+c^-)|\Lambda\rangle=0.
\end{equation}
where the term with $c^+c^-$ is picked up when $c^0$ is anti-commuted through $Q_{cs}$. Due to the ghost-vacuum structure, the equation breaks up into four components
\begin{align}\label{eq:VaryingPhiComponents1}
\delta/\delta A&\Rightarrow G_0A+G_-H=\epsilon\big((G_3-\sigma_+)\widetilde{A}-\sigma_-\widetilde{F}+G_-\widetilde{B}\big),\\\label{eq:VaryingPhiComponents2}
\delta/\delta F&\Rightarrow G_0F+G_+H=\epsilon\big((G_3+\sigma_+)\widetilde{F}+\sigma_-\widetilde{A}+G_+\widetilde{B}\big),\\\label{eq:VaryingPhiComponents3}
\delta/\delta H&\Rightarrow G_+A-G_-F-2H=-\epsilon\widetilde{B},\\\label{eq:VaryingPhiComponents4}
\delta/\delta B&\Rightarrow G_+\widetilde{A}-G_-\widetilde{F}-\widetilde{H}=0.
\end{align}
where it is indicated from which variations of components the equations follow if the component action is worked out from \eqref{eq:ActionLagrangeMultplier}. Then varying with respect to $|\Lambda\rangle$ produces
\begin{equation}\label{eq:VaryingLambda}
(c^0Q_{cs}-c^+c^-)|\Phi\rangle=0.
\end{equation}
Working out the components of this equation yield
\begin{align}\label{eq:VaryingLambdaA}
\delta/\delta\widetilde{A}&\Rightarrow (G_3-\sigma_+)A+G_-B-\sigma_-F=0,\\\label{eq:VaryingLambdaF}
\delta/\delta\widetilde{F}&\Rightarrow (G_3+\sigma_+)F+G_+B+\sigma_-A=0,\\\label{eq:VaryingLambdaB}
\delta/\delta\widetilde{H}&\Rightarrow 2B=0,\\\label{eq:VaryingLambdaH}
\delta/\delta\widetilde{B}&\Rightarrow G_+A-G_-F-H=0.
\end{align}

The system is clearly over-constrained in that equations \eqref{eq:VaryingLambdaB} and \eqref{eq:VaryingLambdaH} together with \eqref{eq:VaryingPhiComponents3} implies $B=0$ and $H=\epsilon\widetilde{B}$. This defect can be remedied by choosing a Lagrange multiplier $|\Lambda\rangle$ with $\widetilde{H}=\widetilde{B}=0$, effectively removing the last two field equations. 

Then defining the Fronsdal matrix operator
\begin{equation}\label{eq:FronsdalMatrix}
\mathbf{F}=
\begin{bmatrix}
G_0 & 0 \\ 0 & G_0
\end{bmatrix}+
\frac{1}{2}\begin{bmatrix}
G_-G_+ & -G_-G_- \\ G_+G_+ & -G_+G_-
\end{bmatrix},
\end{equation}
the field equations \eqref{eq:VaryingPhiComponents1}-\eqref{eq:VaryingPhiComponents3} can be reduced to
\begin{equation}\label{eq:AFFieldEquations}
\mathbf{F}\begin{bmatrix}
A \\ F
\end{bmatrix}=
\epsilon\,\mathbf{T}\begin{bmatrix}
\widetilde A \\ \widetilde F
\end{bmatrix}
\end{equation}
Of the remaining field equations, \eqref{eq:VaryingLambdaA} and \eqref{eq:VaryingLambdaB} can be written (see \eqref{eq:MatrixTraceFieldEquations})
\begin{equation}\label{eq:MatrixTraceFieldEquations2}
\mathbf{T}
\begin{bmatrix}
A \\ F
\end{bmatrix}
=
-\begin{bmatrix}
G_- \\ G_+
\end{bmatrix}
B.
\end{equation}

It is easy check that the operator matrices $\mathbf{F}$ and $\mathbf{T}$ commute. Thus applying $\mathbf{T}$ on both sides of equation \eqref{eq:AFFieldEquations} yields
\begin{equation}
\mathbf{F}\mathbf{T}
\begin{bmatrix}
A \\ F
\end{bmatrix}=
\epsilon\,\mathbf{T}^2\begin{bmatrix}
\widetilde A \\ \widetilde F
\end{bmatrix}
\end{equation}
and using \eqref{eq:MatrixTraceFieldEquations2} we get
\begin{equation}
-\mathbf{F}\begin{bmatrix}
G_- \\ G_+
\end{bmatrix}B
=\epsilon\,\mathbf{T}^2\begin{bmatrix}
\widetilde A \\ \widetilde F
\end{bmatrix}
\end{equation}
However, since the left hand side of this equation is identically zero, we get
\begin{equation}
\mathbf{T}^2\begin{bmatrix}
\widetilde A \\ \widetilde F
\end{bmatrix}=0
\end{equation}
On-shell we can chose for a solution $\widetilde A=\widetilde F=0$ or at least 
\begin{equation}
\mathbf{T}\begin{bmatrix}
\widetilde A \\ \widetilde F
\end{bmatrix}=0
\end{equation}
thus reproducing the contiuous spin field equations of section \ref{subsec:FieldEquations}.
\section{Physical rationale}\label{sec:PhysicalRationale}
There is a  nice physical rationale for the constraint structure. Think of a mechanical system consisting of two point particles with coordinates $t^\mu$ and $b^\mu$ and corresponding canonical momenta $u_\mu$, $d_\mu$. These phase space variables can be called ''end-point'' variables. The center of motion $x^\mu$ and relative coordinates $\xi^\mu$ are defined by
\begin{equation}\label{eq:TopBottomTranscription}
x^\mu={1\over 2}(t^\mu+b^\mu),\quad\xi^\mu={1\over 2}(t^\mu-b^\mu).
\end{equation}
The canonical conjugate momenta $p_\mu$, $\pi_\mu$ are
\begin{equation}\label{eq:UpDownTranscription}
p^\mu=u^\mu+d^\mu,\quad\pi^\mu=u^\mu-d^\mu.
\end{equation}
Consider the two end-points to be moving with the velocity of light so that the relevant constraints are
\begin{equation}\label{eq:EndPointConstraintsMomentum}
u^2\approx0\quad\quad\text{and}\quad\quad d^2\approx0.
\end{equation}
A further natural constraint is to require the center of motion momentum $p$ to have a constant Lorentz-scalar product with the relative coordinate $\xi$ so that
\begin{equation}\label{eq:EndPointConstraintsTransversality}
\xi\cdot p\approx\beta.
\end{equation}
The last constraint is requiring the Lorentz-scalar product of the end-point momenta $u$ and $d$ to be constant
\begin{equation}\label{eq:EndPointConstraintsUD}
u\cdot d\approx \omega^2.
\end{equation}
This last constraint corresponds to $G_3$. It is easy to see that these constraints can be linearly recombined into the constraints used in this paper.\footnote{If one does not want to consider the CS case, the right hand sides of the equations \eqref{eq:EndPointConstraintsTransversality} and \eqref{eq:EndPointConstraintsUD} can be set to zero.}

However, a little bit more can be said, in that the fourth constraint cannot be avoided, it must be included. This is because the three constraints in equations \eqref{eq:EndPointConstraintsMomentum} and \eqref{eq:EndPointConstraintsTransversality} do not form a first class algebra by themselves. One way of seeing this is to linearly recombine the constraints in \eqref{eq:EndPointConstraintsMomentum} into
\begin{equation}
2(u^2+d^2)=p^2+\pi^2\approx 0\,,\quad\quad u^2-d^2=\pi\cdot p\approx0.
\end{equation}
Then, working with Poisson brackets we get
\begin{equation}\label{eq:NonFirstClassAlgebra}
\{p^2+\pi^2,\xi\cdot p\}=-2\pi\cdot p\,,\quad\quad\{\pi\cdot p,\xi\cdot p\}=-p^2,
\end{equation}
which is not first class since we do not have $p^2\approx0$. However, requiring $u\cdot d=\frac{1}{4}(p^2-\pi^2)\approx0$ provides us with the missing $p^2$ constraint. Again we have an algebra with four first class constraints.

This leads to a puzzle regarding the conventional three-constraint higher spin theory. What does it correspond to in terms of end-point coordinates and momenta? Transcribing the set $\{p^2\,,\,\alpha\cdot p\,,\,\alpha^\dagger\cdot p\}$ first to $\{p^2\,,\,\xi\cdot p\,,\,\pi\cdot p\}$ and then to end-point variables, we get the constraints
\begin{equation}
(t-b)\cdot(u+d)\approx0\,,\quad\quad u^2+u\cdot d\approx0\,,\quad\quad d^2+u\cdot d\approx0.
\end{equation}
The interpretation is that in conventional higher spin theory, the end-points of the underlying mechanical model do not move with the velocity of light. To recover that, the trace constraint $u\cdot d$ must again included. What's even more puzzling, the operator $u\cdot d$ plays a still not fully understood role in the interacting theory. It looks like it provides the internal dynamics of the mechanical system.

\section{Concluding remarks}\label{sec:ConcludingRemarks}
Already the triplet BRST-theory of free higher spin gauge fields is free from tracelessness constraints, and thus ''unconstrained''. But it is so at the price of having at least a doublet of fields $(A,F)$ where $F$ propagates lower spin unphysical components. The fields $A$ and $F$ are independent but coupled through the field equations. The objectives of unconstrained formulations are to minimize the number of extra fields needed. The fields $F$ are substituted for by trace-like equations at the price of introducing Lagrange multiplier fields into the action. It seems from the present work that such formulations can be understood, in a BRST framework, as arising from various four-constraint theories. Somewhat surprisingly it seems that a four-constraint theory modeling the continuous spin representations does precisely this, at least in the special case with $\beta=\omega^2=0$ where there is no level mixing. But clearly, more work is needed to sort out the details.

Furthermore, as argued above, there seems to be an underlying physical rationale for the trace constraints. When expressed at the level of Poincar{\'e} covariant fields, it is manifested not through conventional gauge invariance, but instead through these awkward tracelessness constraints. But rather than being just an inconvenience, this very fact may hint at a dynamical principle behind the interactions, based not on the harmonic oscillator equation $\pi^2+\xi^2=0$, but on the $u\cdot d$ equation.

\section{Vacua and ghosts}\label{sec:NotationConventions}
The ghost structure is well known but let us fix notation. Corresponding to the reparameterization constraint $G_0$ we have the pair of Hermitean ghosts $(c_0,b_0)$ with anti-commutator $\{c^0,b_0\}=1$. Their presence leads to a degenerate vacuum with $|-\rangle=c^0|+\rangle$ and $|+\rangle=b_0|-\rangle$. On the other hand, the gauge constraints $G_+$ and $G_-$ go with the conjugate ghost pairs $(c^+,b_+)$ and $(c^-,b_-)$. Their essential properties are summarized in $(c^-)^\dagger=c^+$, $(b_-)^\dagger=b_+$ and the anti-commutators $\{c^+,b_+\}=1$, $\{c^-,b_-\}=1$.

The full ghost complex that results is given in table \ref{tab:HigherSpinGhostComplex}, where $\rm gh_m$ denotes the (mechanical) ghost number. A $|+\rangle$ vacuum is given  ghost number $-1/2$ and the $c$-ghosts and $b$-ghosts have ghost numbers $1$ and $-1$ respectively. The $b_-$ and $c^+$ operators are creators while $b_+$ and $c^-$ are annihilators.

\begin{table}[h]
\centering
\begin{tabular}{c|c|c|c|c|}\hline
$\rm gh_m(\cdot)$ & 3/2 & 1/2 & -1/2 & -3/2 \\ \hline
 & & $c^0|+\rangle$ & $|+\rangle$ & \\ 
 & $c^+c^0|+\rangle$ & $c^+|+\rangle$ & $b_-c^0|+\rangle$ & $b_-|+\rangle$ \\ 
  & & $c^+b_-c^0|+\rangle$ & $c^+b_-|+\rangle$ &  \\  \cline{2-5}
\end{tabular}
\caption{Higher spin ghost complex}
\label{tab:HigherSpinGhostComplex}
\end{table}

The continuous spin ghost complex with the doubly degenerate vacuum structure can then be constructed as in table \ref{tab:FourConstraintSpinGhostComplex}. 

\begin{table}[h]
\centering
\begin{tabular}{c|c|c|c|c|c|}\hline
$\rm gh_m(\cdot)$ & 2 & 1 & 0 & -1 & -2 \\ \hline

 & & $|--\rangle$ & $|+-\rangle,|-+\rangle$ & $|++\rangle$ &\\ 

 & $c^+|--\rangle$ & $c^+|-+\rangle$ & $b_-|--\rangle$ & $b_-|-+\rangle$ & $b_-|++\rangle$\\

 & & $c^+|+-\rangle$ & $c^+b_-|-+\rangle$ & $b_-|+-\rangle$ &\\

 & & $c^+b_-|--\rangle$ & $c^+b_-|+-\rangle$ & $c^+b_-|++\rangle$ &\\

 & & & $c^+|++\rangle$ & &\\  \cline{2-6}
\end{tabular}
\caption{Continuous spin ghost complex}
\label{tab:FourConstraintSpinGhostComplex}
\end{table}

For single degenerate ghost vacua we have $\langle+|-\rangle=\langle-|+\rangle=1$ and $\langle+|+\rangle=\langle-|-\rangle=0$ with the $+$ vacuum even and the $-$ vacuum odd. Consequently we must have
\begin{equation}
\begin{split}
\langle+_3+_0|-_0-_3\rangle&=-\langle+_3|-_3\rangle\langle+_0|-_0\rangle=-1,\\
\langle+_3-_0|+_0-_3\rangle&=-\langle+_3|-_3\rangle\langle-_0|+_0\rangle=-1,\\
\langle-_3+_0|-_0+_3\rangle&=\langle-_3|+_3\rangle\langle+_0|-_0\rangle=1,\\
\langle-_3-_0|+_0+_3\rangle&=\langle-_3|+_3\rangle\langle-_0|+_0\rangle=1.
\end{split}
\end{equation}
All other combinations are zero. When no confusion can arise, the subscripts are dropped. All vacua are ordinary vacua for the bosonic oscillators so that $\alpha_\mu|+_3+_3\rangle=0$ et cetera.

\pagebreak

\end{document}